\documentclass[preprint,showpacs,showkeys,preprintnumbers,amsmath,amssymb]
{revtex4}

\usepackage{graphicx}
\usepackage{dcolumn}
\usepackage{bm}
\usepackage[latin1]{inputenc}

\def\Tr{\mathrm{Tr}}
\def\d{\mathrm{d}}
\def\e{\mathrm{e}}
\def\ket#1{\left| #1\right\rangle}

\begin{document}

\author{R. Rossi Jr.}
\author{M. C. Nemes}
\affiliation{Departamento de F\'{\i}sica, Instituto de Ciências Exatas,
Universidade Federal de Minas Gerais, C.P. 702, 30161-970, Belo Horizonte, MG, Brazil}
\author{J. G. Peixoto de Faria}
\affiliation{Departamento Acadêmico de Ciências Básicas, Centro Federal de Educação
Tecnológica de Minas Gerais, 30510-000, Belo Horizonte, MG, Brazil}

\title{Atomic detection in microwave cavity experiments: a
dynamical model}

\begin{abstract}
We construct a model for the detection of one atom maser in the
context of cavity Quantum Electrodynamics (QED) used to study
coherence properties of superpositions of electromagnetic modes.
Analytic expressions for the atomic ionization are obtained,
considering the imperfections of the measurement process due to
the probabilistic nature of the interactions between the
ionization field and the atoms. Limited efficiency and false
counting rates are considered in a dynamical context, and
consequent results on the information about the state of the
cavity modes are obtained.
\end{abstract}

\maketitle

\section{Introduction}

The quantum interaction between two level Rydberg atoms and one
microwave mode inside a high quality factor (\textit{Q}) cavity has been crucial
for our understanding of dissipation and decoherence in quantum
mechanics \cite{art1,art2}. Usually in cavity 
quantum electrodynamics (QED) experiments,
Rydberg atoms cross an experimental array constituted of two
Ramsey zones and a high \textit{Q} cavity. Thereafter their final state is
detected in two stages. Firstly, they cross an electromagnetic
field which is built to ionize the highest state of the atom. The
second detection zone is designed to detect the lower atomic
level.

Most of the work available in literature about the detection
process is based on statistical assumptions. The pumped
atoms are statistically independent, so that their arrival times
are subject to a Poissonian or other statistics \cite{art3, art4,
art5}. The basic idea is that atoms arrive at random and they are
recorded at equally random times, so the only reproducible data
are statistical. In this context one is lead to studying the
statistic of detector clicks. There are numerical studies
\cite{art6,art7} and also analytical results by Rempe and Walther
\cite{art8} and also by Paul and Richter \cite{art9}.

In the present contribution we propose (to our knowledge for the
first time) a dynamical model for the detection process. 
We assume that the atom undergoes the influence of a classical
electrical field when it traverses the detection zones.
The net effect of this field is to couple one (in the case of
intrinsically inefficient detectors) or two (in the case of 
detectors that register false countings) discrete atomic levels to 
the continuum. If the atom is ionized, a transition to the continuum
has ocurred and a classical signal -- a ``click" -- is generated
in the correspondent detector. However, if the atom remains in
one of the two discrete levels, no click is registered by the
detector. 

Since the atom works as a probe to the field 
stored in a high-\textit{Q} cavity, the click or no-click registered by
the detectors represents a gain of information about the
state of the compound system formed by the atom 
and by the high-\textit{Q} cavity field. Hence,
the process of detection can be divided in two parts. First,
the state of the compound system atom--high-\textit{Q} cavity field
undergoes an unitary evolution during the passage of 
the atom through each detection zone. Next, the resultant state
is projected into a proper subspace defined as follows: if a click is
registered, this subspace is formed by the set of the
states that form the continuum; otherwise, this subspace 
corresponds to the states associated to the two discrete levels.

This paper is organized in the following way. In 
Sec. II, we treat a model for intrinsically inefficient detectors.
The analytical form of our results are exactly the same as those in
reference \cite{art10}, whose derivation is based on statistical
and physically plausible arguments.
In Sec. III, we study the possibility of the detectors 
perform false countings. In this case, we found that the probability
of a click depends on the ``non-diagonal" terms of the 
state of the system atom--high-Q cavity field.
We calculate the fidelity of the field states in high-\textit{Q} cavity
after the measurement process considering the two kinds of
imperfections, limited efficiency and false detections.
Sec. IV contains a summary of the results and conclusions.

\section{Model for Inefficient Detector}

The ionization process of an atom due to its interaction with an
electromagnetic field is considered in a quantum context.
Therefore a finite probability of non-excitation will exist. That
is what we call an intrinsic (i.e., quantum mechanical)
inefficiency.

The Hamiltonian which describes the interaction between two level
atoms and ionization field on the first detection zone, and takes
into account only the intrinsic inefficiency of the process is
given by (the hamiltonian for the second detection zone can been
obtained replacing the index $e$ by $g$):
\begin{equation}
H_{1e}= \epsilon_{e}|e\rangle\langle
e|+\epsilon_{g}|g\rangle\langle g|+\int
dk\epsilon_{k}|k\rangle\langle k|+v_{e}\int dk (|e\rangle\langle
k|+|k\rangle\langle e|) \: .\label{h1}
\end{equation}
The first and second terms in the Hamiltonian stand for 
the two discrete atomic levels $|e\rangle$ and
$|g\rangle$, excited and ground state respectively, with energies
$\epsilon_{e}$ and $\epsilon_{g}$. The third term represents its
continuum spectrum. The last term accounts for the coupling
between the highest discrete level and the continuum. The
strength of this interaction is given by the parameter $v_{e}$,
assumed state independent for simplicity. This term is responsible
for the ionization of the atom. Since we are dealing with a
quantum mechanical process, which is intrinsically probabilistic,
we will also have to consider the possibility of non-ionization of
the atom.

Following Cohen \cite{art11}, the evolution of the discrete state
$|e\rangle$ according to (\ref{h1}) is given by:
\begin{equation}
|\psi(t)\rangle = \e^{-iH_{1e}t/\hbar}|e\rangle = \int d\mu\langle
\psi^{e}_{\mu}|e\rangle
\e^{-i\epsilon^{e}_{\mu}t/\hbar}|\psi_{\mu}\rangle\:,
\end{equation}
where $\{|\psi^{e}_{\mu}\rangle\}$ and $\{\epsilon^{e}_{\mu}\}$
correspond to the set of eigenvectors and eigenvalues of $H_{1e}$
respectively. The coefficients $\langle\psi^{e}_{\mu}|e\rangle$
and $\langle\psi^{e}_{\mu}|k\rangle$ may be written as
\begin{subequations}
    \label{coeff1}
    \begin{align}
        \langle\psi^{e}_{\mu}|e\rangle &= \frac{1}{\left[1+\int
        \d k^{'}\left(\frac{v}{\epsilon^{e}_{\mu}
        -\epsilon_{k^{'}}}\right)^{2}\right]^{1/2}}\:, \label{psi-mu-e}\\
        \langle\psi^{e}_{\mu}|k\rangle &= \frac{v/(\epsilon^{e}_{\mu}
        -\epsilon_{k})}{\left[1+\int
        \d k^{'}\left(\frac{v}{\epsilon^{e}_{\mu}
        -\epsilon_{k^{'}}}\right)^{2}\right]^{1/2}} \: . \label{psi-mu-k}
    \end{align}
\end{subequations}
Accordingly, the ionization probability can be obtained as follows
\begin{eqnarray}
    p_{e}&=&\int \d k |\langle k| \psi(t)\rangle|^{2} \notag \\
    &=&\int \d k \left|\int \d \mu \langle\psi^{e}_{\mu}|e\rangle \langle
    k|\psi^{e}_{\mu}\rangle \e^{-i\epsilon^{e}_{\mu}t/\hbar}
    \right|^{2} \label{ef} \: .
\end{eqnarray}
This probability defines the detector's efficiency. Therefore the
non-detection probability is given by $1-p_{e}=\left|\int \d  \mu
\e^{-i\epsilon^{e}_{\mu}t/\hbar}|\langle\psi^{e}_{\mu}|e\rangle|^{2}
\right|^{2}$. After some simplifications (see \cite{art11}) the
non-detection probability can be written as
\begin{equation}
    1-p_{e}=\e^{-\Gamma|t|} \: ,
\end{equation}
where $\Gamma$ is the transition rate from discrete to the
continuum level, calculated from Fermi's golden rule.
$\Gamma$ is given by
\begin{equation}
    \Gamma= \frac{2\pi \rho(E)}{\hbar} \:,
\end{equation}
where $\rho(E)$ is the level density per unity energy.
In the limit where the atom ionization time can be considered to
be infinite (in some experimental context) we will have a perfect
detector.

Following the same procedure for $H_{1g}$ we find $p_{g}=\int \d k
\left|\int \d \mu \langle\psi^{g}_{\mu}|g\rangle \langle
k|\psi^{g}_{\mu}\rangle \e^{-i\epsilon^{g}_{\mu}t/\hbar}
\right|^{2}$, where $\{|\psi^{g}_{\mu}\rangle\}$ and
$\{\epsilon^{g}_{\mu}\}$ correspond to the set of eigenvectors and
eigenvalues of $H_{1g}$ respectively.

\begin{figure}
	\centering
	\includegraphics[scale=0.5]{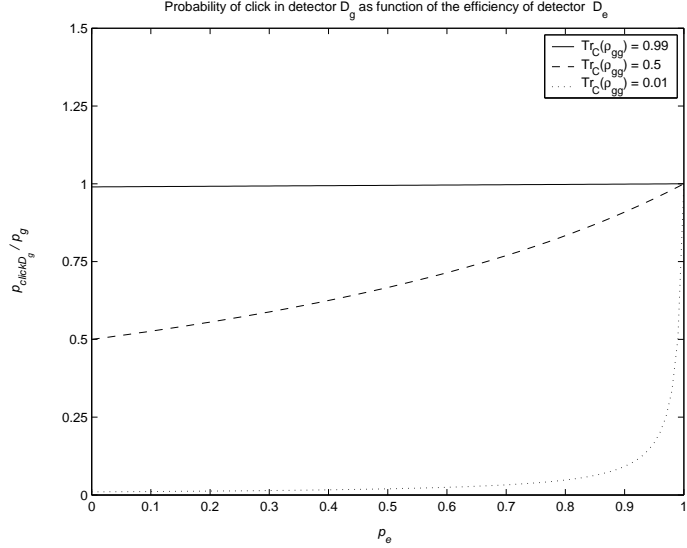}
	\caption{Influence of the efficiency of the detector $D_e$ ($p_e$) on
	the ``normalized" probability of click in the detector $D_g$ 
	($p_{\mathrm{click}D_g}/p_g$), 
	for different values of $\Tr_C\left(\rho_{gg}\right)$. The efficiency of the detector
	$D_g$ ($p_g$) just limits the maximum value reached by $p_{\mathrm{click}D_g}$
	and does not modify its qualitative behavior as function of $p_e$.
	}
	\label{fig1}
\end{figure}

\subsection{An exemple: cavity QED }

As an exemple of applicability of the model, let us study the
interactions between two level atoms and their detection through
ionization fields in cavity QED experiments. The state of the
system atom--high-\textit{Q} cavity field can be written as
\begin{equation}
    \rho_{AC}(0)=\rho_{ee}|e\rangle\langle e|+
    \rho_{eg}|e\rangle\langle g|+\rho_{ge}|g\rangle\langle
    e|+\rho_{gg}|g\rangle\langle g| \: . \label{inicial}
\end{equation}
This state represents the most general state (in the system
atom-cavity field) immediately before the interaction between the
atom and the detectors. The symbols $\rho_{ee}$, $\rho_{eg}$, $\rho_{ge}$ and
$\rho_{gg}$ are operators in the cavity field subsystem.

The interaction between the atom and the first detection
zone ($D_{e}$) can be separated in two steps. Firstly, a quantum
unitary evolution governed by the Hamiltonian $H_{1e}$ given by Eq. (\ref{h1})
during the time interval $t_{1}$. The atom-cavity field state, after this
process, is given by
\begin{equation}
    \rho_{AC}(t_{1})=\e^{-iH_{1e}t_{1}/\hbar}\rho_{AC}(0)\e^{iH_{1e}t_{1}/\hbar}\: .
\end{equation}

Now, in the second step, at the time $t_{1}$, a classical signal is
generated. If the detector clicks, we will know that the atom was
ionized, so $\rho_{AC}(t_{1})$ must be projected into the subspace
$\{|k\rangle\}$. Although, if $D_{e}$ does not click we know that
the atomic state must be projected into subspace spanned by the
discrete levels $\{|e\rangle,
|g\rangle\}$. The maximum value that $t_{1}$ can assume is
$t_{1}'$ which is the time taken by the atom to cross $D_{e}$
completely. So, up to $t_{1}'$, we will certainly acquire
information about the system. This revealed information plays an
essencial role into $\rho_{AC}$'s evolution. So we are aware that
before the interaction with $D_{g}$, the state $\rho_{AC}(t_{1})$ must be
projected into properly subspace.

We can calculate the probability of a click in $D_{e}$
\begin{equation}
    p_{\mathrm{click}D_{e}}=\int \d k \Tr\left(|k\rangle\langle
    k|\rho_{AC}(t_{1})\right)=\int \d k \Tr\left(|k\rangle\langle
    k|\rho_{ee}\e^{-iH_{1e}t_{1}/\hbar}|e\rangle\langle
    e|\e^{iH_{1e}t_{1}/\hbar}\right).
\end{equation}
Then, using Eq. (\ref{ef}), we may write
\begin{equation}
    p_{\mathrm{click} D_{e}}=p_{e}\Tr_{C}(\rho_{ee}) \: ,
\end{equation}
where $\Tr_{C}$ is the partial trace on the cavity-field subspace.
This product can be interpreted as the efficiency of $D_{e}$
($p_{e}$) times the probability of click on a perfect detector
after the interaction with the state $\rho_{AC}(0)$
($\Tr_{C}(\rho_{ee})$).

The non-click probability is:
\begin{equation}
    p_{\mathrm{non-click} D_{e}}=\Tr\left[\left(|e\rangle\langle e| +
    |g\rangle\langle
    g|\right)\rho_{AC}(t_{1})\right]=\Tr_{C}(\rho_{ee})+
    \Tr_{C}(\rho_{gg}) - p_{e}\Tr_{C}(\rho_{ee})\: .
\end{equation}
From the
normalization of $\rho_{AC}(0)$, $\Tr_{C}(\rho_{ee})+ Tr_{C}(\rho_{gg}) = 1$,
we can write
\begin{equation}
    p_{\mathrm{non-click} D_{e}}= 1 - p_{e}\Tr_{C}(\rho_{ee}) \: .
\end{equation}

After the non-click stage the system is in the state:
\begin{eqnarray}
    \rho_{\mathrm{non-click}}(t_{1})&=& \frac{\left(|e\rangle\langle e| +
    |g\rangle\langle g|\right)\rho_{AC}(t_{1})\left(|e\rangle\langle
    e| + |g\rangle\langle g|\right)}{N}\\
    &=& \frac{\rho_{gg}|g\rangle\langle
    g|+\rho_{ee}(1-p_{e})|e\rangle\langle
    e|+(\rho_{eg}\e^{i\epsilon_{g}t_{1}/\hbar}\int \d\mu
    e^{-i\epsilon_{\mu}t_{1}/\hbar}|\langle\psi_{\mu} |e\rangle|^{2}
    |e\rangle\langle g|+h.c.)}{N}   \: ,\notag
\end{eqnarray}
where $N=1-p_{e}\Tr_{c}\rho_{ee}$. Note that if the efficiency is equal to
unity $(p_{e}=1)$, the reduced state operator on atomic subspace can be
written as $\rho_{gg}|g\rangle\langle g|$. This result was
expected, as we know, for perfect detectors a non-click in $D_{e}$
would lead to the projection $|g\rangle\langle
g|\rho_{AC}(0)|g\rangle\langle g|$.

When the atom is not ionized on $D_{e}$, it continues the journey
and passes through the second detection zone ($D_{g}$). Let us set
the interaction time between atom and the electromagnetic field
inside $D_{g}$ by $t_{2}$. The temporal evolution that models
this interaction is again unitary:
\begin{equation}
    \rho_{AC}(t_{2})=\frac{1}{N}\e^{-iH_{1g}(t_{2}-t_{1})\hbar}
    \rho_{\mathrm{non-click}}(t_{1})\e^{iH_{1g}(t_{2}-t_{1})\hbar}\: .
\end{equation}

So, the probability of click in $D_{g}$ is
\begin{equation}
    p_{\mathrm{click} D_{g}}=
    \frac{p_{g}\Tr_{C}(\rho_{gg})}{1-p_{e}\Tr_{C}(\rho_{ee})}\:,
\end{equation}
Note that this probability depends on the efficiency of the first
detector $(p_{e})$. Now let us examine some limits. For $p_{e}=0$,
this is equivalent to the situation where the first detector is
absent, so the atom interacts just with the second ionization zone
$D_{g}$ and the probability of click is $p_{g}\Tr_{C}(\rho_{gg})$,
as expected. If $p_{e}=1$, the first detector is perfect, therefore
we know that when an atom crosses $D_{e}$ without been detected, the
system goes to the state $|g\rangle$, as discussed before, and the
probability of click is the efficiency of $D_{g}$ ($p_{g}$). If both
detectors are perfect $(p_{e}=p_{g}=1)$, $p_{\mathrm{click}
D_{g}}=1$ because the second detector will not miss any atom
prepared in $|g\rangle$.

A more complete analysis of $p_{\mathrm{click} D_{g}}$ for different
$\rho_{AC}(0)$ is shown in Fig. 1. The behavior of the curves
associate to $\Tr_{C}(\rho_{gg})=0.5$ and  $\Tr_{C}(\rho_{gg})=0.01$
reflects the fact that a non-click on a very efficient $D_{e}$
($p_{e}\approx 1$) raises the probability $p_{\mathrm{click}
D_{g}}$, even if the atom is practically prepared in the state
$|e\rangle\langle e|$ ($\Tr_{C}(\rho_{gg})=0.01$). On the other
hand, if the atom is practically prepared in the state
$|g\rangle\langle g|$ ($\Tr_{C}(\rho_{gg})=0.99$), $p_{e}$ does not
affect $p_{\mathrm{click} D_{g}}$.

The probability of non-click in $D_{g}$ is
\begin{equation}
    1-p_{\mathrm{click}
    D_{g}}=\frac{1-p_{e}\Tr_{C}\rho_{ee}-p_{g}\Tr_{C}\rho_{gg}}{1-p_{e}\Tr_{C}\rho_{ee}}
\end{equation}
When the atom crosses both detectors without being detected, it
reduces the field state inside the cavity to
\begin{equation}
    \rho_{C}'=\frac{\Tr_{A}\left[\left(|e\rangle\langle e| +
    |g\rangle\langle
    g|\right)\rho_{AC}(t_{2}-t_{1})\left(|e\rangle\langle e| +
    |g\rangle\langle
    g|\right)\right]}{\Tr\left[\left(|e\rangle\langle e| +
    |g\rangle\langle
    g|\right)\rho_{AC}(t_{2}-t_{1})\left(|e\rangle\langle e| +
    |g\rangle\langle g|\right)\right]}\: .
\end{equation}
Here, $\Tr_{A} $ stands for the trace in the atomic variables.
Now, using the definition (\ref{ef}) we may write:
\begin{equation}
    \rho_{C}'=\frac{(1-p_{e})\rho_{ee}+(1-p_{g})\rho_{gg}}
    {1-p_{e}\Tr_{C}(\rho_{ee})-p_{g}\Tr_{C}(\rho_{gg})}\: .\label{apos}
\end{equation}
The form of this result is in complete agreement with the one in
\cite{art10}, where the authors used statistics arguments to
derive the expression (\ref{apos}).

\section{Model for false detections}

In addition to the previous intrinsically inefficient detector we
extend the model to include false detections. The hamiltonian for
the first detection zone $D_{e}$ is given by (the hamiltonian for
the second detection zone can been obtained replacing the index
$e$ by $g$):
\begin{eqnarray}
    H_{2e}&=&\epsilon_{e}|e\rangle\langle
    e|+\epsilon_{g}|g\rangle\langle g|+\int
    \d k\epsilon_{k}|k\rangle\langle
    k| \label{h2}\\
    &+&w_{e}\int \d k (|e\rangle\langle k|+|k\rangle\langle e|)+
    w_{g}\int \d k (|g\rangle\langle k|+|k\rangle\langle g|)\: , \notag
\end{eqnarray}
where $w_{e}$ and $w_{g}$ are real coupling constants. The second
interaction term (the last one in the equation above) represents
the coupling between $|g\rangle$ and the continuum, so it is
responsible for wrong detections. On the other hand, the previous
one is responsible for the correct ones.

For simplicity we are going to define the complex coefficient
\begin{equation}
    q^{e}_{a,b}=\int \d \eta \e^{-i\epsilon^{e}_{\eta}t/\hbar}\langle
    \phi^{e}_{\eta}|a\rangle \langle b|\phi^{e}_{\eta}\rangle \: ,
    \label{co}
\end{equation}
where $\{|\phi^{e}_{\eta}\rangle\}$ and $\{\epsilon^{e}_{\eta}\}$
are the set of eigenvectors and eigenvalues of $H_{2e}$
respectively. $a$ and $b$ are indexes that may represent continuum
and discrete eigenvectors. The explicit form of the coefficients
inside the integral and relative discussions are in the appendix.
The notation is as follows: the upper index indicates which detection zone
the atom is passing through. The two lower indexes, $a$ and $b$, represent its
initial incoming state and its final state after traversing the
detector, respectively.
One can notice that $\int \d k |q^{e}_{e,k}|^{2}$ is the probability
of an atom prepared in $|e\rangle$ to be ionized inside $D_{e}$,
this can be understood as the efficiency of $D_{e}$. $\int \d k
|q^{e}_{g,k}|^{2}$ is the probability of an atom prepared in
$|g\rangle$ to be ionized inside $D_{e}$, i.e., the probability of
a wrong detection.

We can see, from (\ref{h2}), that the unitary evolution of this
system allows for an indirect coupling between the two discrete
levels. So we can take into account $|q^{e}_{e,g}|^{2}$ which is
the probability of a transition between the two discrete levels.
$|q^{e}_{e,e}|^{2}$ ($|q^{e}_{g,g}|^{2}$) is the probability of an
atom prepared in $|e\rangle$ ($|g\rangle$) to interact with the
electromagnetic field inside $D_{e}$ and do not change level. We
can also notice that $\int \d k|q^{g}_{e,k}|^{2}$ ($\int \d k |q^{g}_{g,k}|^{2}$)
is the probability that an atom prepared in $|e\rangle$ ($|g\rangle$)
to be ionized inside $D_{g}$.

\subsection{An example: cavity QED}

As we did for the intrinsically inefficient detectors, the
interaction between atoms and false counting detectors can be
separated in two processes: firstly, an unitary evolution of the
initial state operator, generated by $H_{2e}$
($H_{2g}$) where $H_{2e}$ ($H_{2g}$) have the
form shown in Eq. (\ref{h2}), and then a projection in a subspace, which
represents the classical information, click or non-click, on the
detector.

Starting from the initial state given by Eq. (\ref{inicial}), and using
the definitions in Eq. (\ref{co}), the probability of click in $D_{e}$ can
be written as
\begin{equation}
    p_{\mathrm{click} D_{e}}=\int \d k |q^{e}_{e,k}|^{2}\Tr_{C}(\rho_{ee}) + \int
    dk |q^{e}_{g,k}|^{2}\Tr_{C}(\rho_{gg}) + \left(\int \d k
    q^{e*}_{e,k}q^{e}_{g,k}\Tr_{C}(\rho_{eg}) + \mathrm{h.c.}\right) \:.
    \label{prob2}
\end{equation}
This expression shows us that $p_{\mathrm{click} D_{e}}$ is sensitive to
interference terms $(\rho_{eg})$ and $(\rho_{ge})$.
If we calculate the value of $p_{\mathrm{click} D_{e}}$ for the initial
state $\rho_{AC}(0)=\rho_{ee}|e\rangle\langle
e|+\rho_{gg}|g\rangle\langle g|$, the answer would be different
from (\ref{prob2}). However, if we do the same, but replacing the
false counting detectors by inefficient or perfect detectors, the
calculated probability would be the same for the two different
initial states. That is due to the fact that this case is
insensitive to interference terms.

In order to compare the modifications on the cavity field due to
atomic interaction with inefficient detectors and false counting
detectors, we calculate the fidelity of the different state
operators. Fidelity between the states $\rho_{A}$ and $\rho_{B}$
measures the overlap between them and is given by
\begin{equation}
    F(\rho_{A},\rho_{B})=\left(\Tr\sqrt{\rho_{A}^{1/2}\rho_{B}\rho_{A}^{1/2}}\right)^{2}
    \: .
\end{equation}

Firstly, let us calcule the fidelity between state operators
$\rho_{A}^{e}$ (which describe the system after the atomic
ionization inside the first detection zone of inefficient
detectors), and $\rho_{B}^{e}$ (which describe the system after
the atomic ionization inside the first detection zone of false
counting detectors). For simplicity, assume that the system atom--high-\textit{Q} cavity
field is found in the following entangled state just before the atom 
reaches the detection zones:
\begin{equation}
    \rho_{AC}(0)=\frac{1}{2}\left(|e,0\rangle\langle
    e,0|+|e,0\rangle\langle g,1|+|g,1\rangle\langle
    e,0|+|g,1\rangle\langle g,1|\right) \: .\label{inicial2}
\end{equation}

After an unitary evolution and the projection on the continuum
subspace, $\rho_{A}^{e}$ and $\rho_{B}^{e}$ can be written as
\begin{subequations}
    \label{roAB}
    \begin{align}
        \rho_{A}^{e}=&|0\rangle\langle 0|\: , \label{roA} \\
        \rho_{B}^{e}=&\frac{\int \d k \left(|q^{e}_{e,k}|^{2}|0\rangle\langle
        0|+|q^{e}_{g,k}|^{2}|1\rangle\langle
        1|+q^{e*}_{g,k}q^{e}_{e,k}|1\rangle\langle
        0|+q^{e*}_{e,k}q^{e}_{g,k}|0\rangle\langle 1|\right)}{\int \d k
        \left(|q^{e}_{e,k}|^{2}+|q^{e}_{g,k}|^{2}\right)} \: , \label{roB}
    \end{align}
\end{subequations}
and the fidelity
\begin{equation}
    F(\rho_{A}^{e},\rho_{B}^{e})=\frac{\int \d k
    |q^{e}_{e,k}|^{2}}{\int \d k\left(|q^{e}_{e,k}|^{2} +
    |q^{e}_{g,k}|^{2}\right)} \: .
\end{equation}
Notice that if the wrong detection probability goes to zero $(\int
dk|q^{e}_{g,k}|^{2}\rightarrow 0)$, the fidelity goes to one,
$F(\rho_{A}^{e},\rho_{B}^{e})\rightarrow 1$, so $\rho_{A}^{e}$
and $\rho_{B}^{e}$ became identical.

Now, we are going to calcule the fidelity between state operators
$\rho_{A}^{g}$, which describes the system after the atomic
ionization inside the second detection zone of inefficient
detectors, and $\rho_{B}^{g}$, which describes the system the
atomic ionization inside the second detection zone of false
counting detectors.
The calculation follows as this: interaction of atom 
with the first detection zone $D_e$, modeled by unitary evolution of
the state given by Eq. \eqref{inicial2} and 
projection on the discrete subspace. Then, the interaction with the
second detection zone $D_g$, modeled again by unitary evolution of the
resultant state and
projection, but now on the continuum subspace. After this, we can
write the fidelity as
\begin{equation}
    F(\rho_{A}^{g},\rho_{B}^{g})=\frac{1}{A}\int
    \d k\left(|q^{e*}_{g,g}|^{2}|q^{g*}_{g,k}|^{2}+
    q^{e*}_{g,g}q^{e}_{g,e}q^{g*}_{g,k}q^{g}_{e,k}+q^{e}_{g,g}q^{e*}_{g,e}q^{g}_{g,k}q^{g*}_{e,k}+
    |q^{e}_{g,e}|^{2}|q^{g}_{e,k}|^{2}\right) \: ,
\end{equation}
where
\begin{eqnarray}
    A&=&\int \d k\left(|q^{e}_{g,e}|^{2}|q^{g}_{g,k}|^{2}+
    q^{e*}_{g,e}q^{e}_{e,e}q^{g*}_{g,k}q^{g}_{e,k}+q^{e}_{g,e}q^{e*}_{e,e}q^{g}_{g,k}q^{g*}_{e,k}+
    |q^{e}_{e,e}|^{2}|q^{g}_{e,k}|^{2}\right) \notag \\
    &+&\int \d k\left(|q^{e}_{g,g}|^{2}|q^{g}_{g,k}|^{2}+
    q^{e*}_{g,g}q^{e}_{g,e}q^{g*}_{g,k}q^{g}_{e,k}+q^{e}_{g,g}q^{e*}_{g,e}q^{g}_{g,k}q^{g*}_{e,k}+
    |q^{e}_{g,e}|^{2}|q^{g}_{e,k}|^{2}\right) \: .
\end{eqnarray}

As we are considering that any transition from a discrete state to
the continuum generates a classical signal, we must not admit the
possibility of indirect coupling between $|e\rangle$ and $|g\rangle$
mediated by the continuum. Therefore, we must assume that
$|q^{e}_{g,e}|^{2}=0$ and we may write:
\begin{equation}
    F(\rho_{A}^{g},\rho_{B}^{g})=\frac{\int
    \d k|q^{e}_{g,g}|^{2}|q^{g}_{g,k}|^{2}}{\int
    \d k\left(|q^{e}_{g,g}|^{2}
    |q^{g}_{g,k}|^{2}+|q^{e}_{e,e}|^{2}|q^{g}_{e,k}|^{2}\right)}.
\end{equation}
If the wrong detections probability in $D_{g}$ goes to zero $(\int
dk|q^{g}_{e,k}|^{2}\rightarrow 0)$ the fidelity goes to one
[$F(\rho_{A}^{g},\rho_{B}^{g})\rightarrow 1$].

\section{Conclusions}

We have presented a dynamical model for the detection process of
atomic levels on field ionization detectors. On the context of
cavity QED, the model allows us to calculate the reduced state
operator, for the field inside the cavity, after the classical
signal generated by the detectors.

The detailed analysis of the detection process also let us introduce
naturally the effects of realistic features of the detectores (e.g.
efficiency and false counting rates) on the study of microwave
cavity experiments. For intrinsically inefficient detectors, we found 
that the probability of a click in the second detection zone
is sensitive to the efficiency of the first one. Besides, our results
are in complete agreement with those obtained in Ref. \cite{art10} 
by different methods.

If one allows the detectors to register false countings, the 
probability of a click is sensitive to the ``non-diagonal" or coherence terms
of the state of the system atom--high-\textit{Q} cavity field.
In fact, false countings are a consequence of the coupling 
between the two discrete atomic levels to the continuum 
in each detection
zone. As a result of this coupling, a click registered in any
detector does not
provide an \textit{unequivocal} information about the
atomic state. The detectors acts as
a ``beam splitter", mixing the two ``paths" $e$ and $g$,
and, to some degree, these ``paths" become undistinguishable.

\appendix*
\section{Evaluation of the coefficients $\left\langle a|\phi^e_\eta\right\rangle$ }
In order to calculate the coefficients inside the integral in \eqref{co},
let us write the eigenvalue equation for $H_{2e}$:
\begin{equation}
    H_{2e}|\phi^{e}_{\eta}\rangle =
    \left[H_{e(0)}+H_{2e(I)}+H_{2e(II)}\right]|\phi^{e}_{\eta}\rangle =
    \epsilon^{e}_{\eta}|\phi^{e}_{\eta}\rangle \:,
    \label{estadoestacionario}
\end{equation}
where $|\phi^{e}_{\eta}\rangle$ and $\epsilon^{e}_{\eta}$ are
eigenvectors and eigenvalues of the Hamiltonian $H_{2e}$,
$H_{e(0)}=\epsilon_{e}|e\rangle\langle
e|+\epsilon_{g}|g\rangle\langle g|+\int
\d k\epsilon_{k}|k\rangle\langle k|$, $H_{2e(I)}=w_e\int \d k
(|e\rangle\langle k|+|k\rangle\langle e|)$ and
$H_{2e(II)}=w_{g}\int \d k (|g\rangle\langle k|+|k\rangle\langle
g|)$.

Let us project equation (\ref{estadoestacionario}) onto
$\ket{ k}$:
\begin{eqnarray}
    \langle k|H_{2e}|\phi^{e}_{\eta}\rangle &=& \langle
    k|H_{e(0)}|\phi^{e}_{\eta}\rangle+\langle
    k|H_{2e(I)}|\phi^{e}_{\eta}\rangle+\langle
    k|H_{2e(II)}|\phi^{e}_{\eta}\rangle \notag \\
    &=& \epsilon_{k}\langle
    k|\phi^{e}_{\eta}\rangle+ w_{e}\langle
    e|\phi^{e}_{\eta}\rangle+w_{g}\langle
    g|\phi^{e}_{\eta}\rangle=\epsilon^{e}_{\eta}\langle
    k|\phi^{e}_{\eta}\rangle \: .\label{k}
\end{eqnarray}
We can do the same with the discrete states $\ket{g}$ and
$\ket{e}$:
\begin{subequations}
    \begin{align}
        \langle g|H_{2e}|\phi^{e}_{\eta}\rangle =&\langle
        g|H_{e(0)}|\phi^{e}_{\eta}\rangle+\langle
        g|H_{2e(I)}|\phi^{e}_{\eta}\rangle+\langle
        g|H_{2e(II)}|\phi^{e}_{\eta}\rangle \notag \\
        =& \epsilon_{g}\langle
        g|\phi^{e}_{\eta}\rangle+w_{g}\int \d k\langle
        k|\phi^{e}_{\eta}\rangle=\epsilon^{e}_{\eta}\langle
        g|\phi^{e}_{\eta}\rangle  \:,\label{f} \\
        \langle e|H_{2e}|\phi^{e}_{\eta}\rangle =&\langle
        e|H_{e(0)}|\phi^{e}_{\eta}\rangle+\langle
        e|H_{2e(I)}|\phi^{e}_{\eta}\rangle+\langle
        e|H_{2e(II)}|\phi^{e}_{\eta}\rangle \notag \\
        =& \epsilon_{e}\langle
        e|\phi^{e}_{\eta}\rangle+w_{e}\int \d k\langle
        k|\phi^{e}_{\eta}\rangle=\epsilon^{e}_{\eta}\langle
        e|\phi^{e}_{\eta}\rangle \:.\label{i}
    \end{align}
\end{subequations}
From the eigenvector's normalization, we can also get the
following expression
\begin{equation}
    |\langle g|\phi^{e}_{\eta}\rangle|^{2}+|\langle
    e|\phi^{e}_{\eta}\rangle|^{2}+\int \d k|\langle
    k|\phi^{e}_{\eta}\rangle|^{2} =1 \:.
\end{equation}

Defining the fundamental energy level as $\epsilon_{g}=0$, and
using Eq. (\ref{f}) and (\ref{i}) we can write
\begin{equation}
    \langle e|\phi^{e}_{\eta}\rangle=
    \frac{\epsilon^{e}_{\eta}w_{e}}{w_{g}(\epsilon^{e}_{\eta}-\epsilon_{e})}\langle
    g|\phi^{e}_{\eta}\rangle \:. \label{final1}
\end{equation}
From Eq. \eqref{k}, we have
\begin{equation}
    \langle k|\phi^{e}_{\eta}\rangle
    =\frac{1}{\epsilon^{e}_{\eta}-\epsilon_{k}}
    \left(w_{g}+\frac{\epsilon^{e}_{\eta}w_{e}^{2}}
    {w_{g}(\epsilon^{e}_{\eta}-\epsilon_{e})}\right)\langle
    g|\phi^{e}_{\eta}\rangle \:,\label{final2}
\end{equation}
and from the normalization condition, we obtain
\begin{equation}
    \langle g|\phi^{e}_{\eta}\rangle=
    \left\{\frac{1}{1+\left[\frac{\epsilon^{e}_{\eta}w_{e}}{w_{g}(\epsilon^{e}_{\eta}-\epsilon_{e})
    }\right]^{2}+\left[w_{g}+\frac{\epsilon^{e}_{\eta}w_{e}^{2}}
    {w_{g}(\epsilon^{e}_{\eta}-\epsilon_{e})}\right]^{2}\int
    \d k\left(\frac{1}{\epsilon^{e}_{\eta}-\epsilon_{k}}\right)^{2}}\right\}^{1/2}
    \:.\label{final3}
\end{equation}
Therefore, Eq. (\ref{final1}), (\ref{final2}) and (\ref{final3}) give
us the explicit form of the three coeficientes.

\end{document}